\documentclass[reprint,aps,prapplied,superscriptaddress,nofootinbib,floatfix]{revtex4-2}

\usepackage{graphicx}
\usepackage[table]{xcolor}   % needed by the comparison table (Fig. 6)
\usepackage{amsmath}
\usepackage[colorlinks=true,allcolors=blue]{hyperref}

\usepackage{comment} %PN added to comment out supp. material

\setcitestyle{super}
\newcommand*{\urlprefix}{}

\newcommand{\IBMSJ}{IBM Research, San Jose, CA, USA}
\newcommand{\IBMZH}{IBM Research, Zurich, Switzerland}
\newcommand{\MuNovi}{Micron Technology, Novi, MI, USA}
\newcommand{\MuFolsom}{Micron Technology, Folsom, CA, USA}
\newcommand{\MuBoise}{Micron Technology, Boise, ID, USA}
\newcommand{\IBMYKT}{IBM Research, Yorktown Heights, NY, USA}
\newcommand{\Patras}{University of Patras, Patras, Greece}
\newcommand{\IBMTKY}{IBM Research, Tokyo, Japan}
\newcommand{\MuTokyo}{Micron Technology, Tokyo, Japan}
\newcommand{\MuSJ}{Micron Technology, San Jose, CA, USA}
\newif\ifvenuepublic \venuepublictrue
\newif\ifdoiknown    \doiknownfalse
\newcommand{\paperdoi}{10.1109/ISSCC.2027.XXXXXXX}  % placeholder until the IEEE Xplore DOI is assigned

\begin{document}

\title{3D-NAND-CIM: A Hybrid-Bonded 3D-NAND-Flash-based Analog In-Memory-Compute macro offering 4.8M~weights/mm$^2$}

\author{Pritish Narayanan}\thanks{These authors contributed equally to this work.}\affiliation{\IBMSJ}
\author{Athanasios Vasilopoulos}\thanks{These authors contributed equally to this work.}\affiliation{\IBMZH}
\author{Geoffrey W. Burr}\affiliation{\IBMSJ}
\author{Manuel Le Gallo}\affiliation{\IBMZH}
\author{Bill Filipiak}\affiliation{\MuNovi}
\author{Hernan Castro}\affiliation{\MuFolsom}
\author{Jeremy Binfet}\affiliation{\MuBoise}
\author{Jeremy Hirst}\affiliation{\MuFolsom}
\author{Henry Utomo}\affiliation{\IBMYKT}
\author{Kunal Parekh}\affiliation{\MuBoise}
\author{Bharat Bhushan}\affiliation{\MuBoise}
\author{Rajasekhar Venigalla}\affiliation{\MuBoise}
\author{Kyle Kirby}\affiliation{\MuBoise}
\author{Urs Egger}\affiliation{\IBMZH}
\author{Jose Luquin}\affiliation{\IBMSJ}
\author{Vara Prasad Jonnalagadda}\affiliation{\IBMZH}
\author{Athanasios Siampakos}\affiliation{\Patras}
\author{Theodore Antonakopoulos}\affiliation{\Patras}
\author{Kohji Hosokawa}\affiliation{\IBMTKY}
\author{Masatoshi Ishii}\affiliation{\IBMTKY}
\author{Yasuteru Kohda}\affiliation{\IBMTKY}
\author{Seiji Munetoh}\affiliation{\IBMTKY}
\author{Abhairaj Singh}\affiliation{\IBMZH}
\author{Alexander Friz}\affiliation{\IBMSJ}
\author{Julian B\"{u}chel}\affiliation{\IBMZH}
\author{Iason Chalas}\affiliation{\IBMZH}
\author{Charles Mackin}\affiliation{\IBMSJ}
\author{Akul Malhotra}\affiliation{\IBMSJ}
\author{Akira Goda}\affiliation{\MuTokyo}
\author{Duane Mills}\affiliation{\MuFolsom}
\author{Gregory Atwood}\affiliation{\MuSJ}
\author{Mark Helm}\affiliation{\MuSJ}
\author{Jeffrey Burns}\affiliation{\IBMYKT}
\author{Vijay Narayanan}\affiliation{\IBMYKT}
\author{Hsinyu Tsai}\affiliation{\IBMSJ}
\author{Abu Sebastian}\affiliation{\IBMZH}

\begin{abstract}
We describe an analog-in-memory-compute macro hybrid-bonding a 14nm logic die onto a 3D NAND Flash memory wafer, achieving weight density of 4.8M weights/mm$^2$. Weights are programmed to 1-of-16 conductances. MACs with current-controlled oscillator ADCs show weight-equivalent precision between 3 and 4 bits for 8-bit inputs on multiple NAND tiers at energy efficiencies up to 4.2 TOPS/W. A charge-retention circuit improves accuracy, with $\sim$0.4\% (3.23\%) accuracy drop on a WideResNet for CIFAR-10 (100).
\end{abstract}

\maketitle

% ---- IEEE copyright notice (required on the first page of the posted accepted version)
\begingroup
\renewcommand{\thefootnote}{}
\footnotetext{%
\ifvenuepublic Accepted at the 2027 IEEE International Solid-State Circuits Conference (ISSCC~2027), San Francisco, CA, USA, as \mbox{Paper~15.5} in Session~15 (Compute-in-Memory), February~16, 2027.\ \fi
\ifdoiknown Published version: \href{https://doi.org/\paperdoi}{doi:\paperdoi}.\ \fi
\textcopyright~2027 IEEE. Personal use of this material is permitted. Permission from IEEE must be obtained for all other uses, in any current or future media, including reprinting/republishing this material for advertising or promotional purposes, creating new collective works, for resale or redistribution to servers or lists, or reuse of any copyrighted component of this work in other works.}
\endgroup

% ============================================================================
%  MAIN TEXT
% ============================================================================

The energy efficiency challenges of conventional general-purpose processors, such as GPUs or CPUs, are becoming a problem as increasingly-capable Artificial Intelligence (AI) models grow to many billions of parameters\cite{Y2026maComputer}.
Analog In-Memory-Compute (AIMC)\cite{HERMES_paper:2022,ARES_paper:2023,TVLSI_paper:2023,Mythic_paper:2022,Y2025singhNatElec} can provide better energy efficiency by performing matrix--vector multiplications in parallel on ``Tiles'' (arrays) of Non-Volatile Memory (NVM) devices.
This removes static weight data movement, and fits well with the recent trend of specialized inference accelerators for the large fully-connected layers within Large Language Models\cite{zhong_disaggregated_inference:2024,NVIDIA_Groq:2026}, preserving precious DRAM capacity for KV-cache.
Up until the present work, AIMC had not demonstrated a path to the very large weight capacities needed for modern AI models, while maintaining its high energy efficiency, high throughput and low latency benefits. To deliver this, we describe 3D-NAND-CIM, a 3D NAND Flash-based AIMC macro created by hybrid-bonding a 14nm logic die containing In-Memory-Compute (IMC) circuitry such as input-drivers and analog-to-digital converters (ADCs) onto a NAND Flash memory wafer.

\begin{figure}[tb]
\centering
\includegraphics[width=\columnwidth]{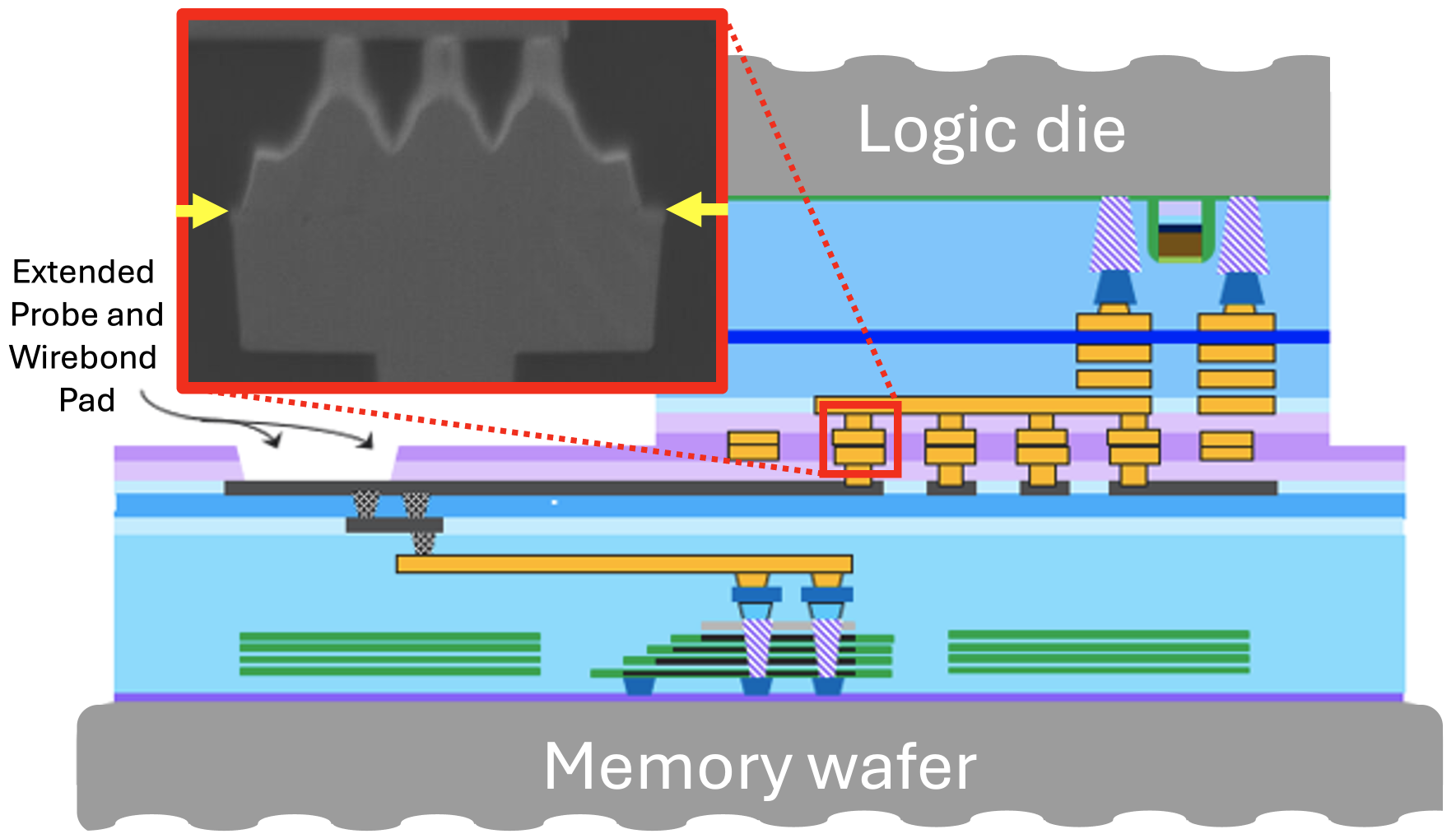}
\caption{3D-NAND-CIM concept vehicle: a 14nm FinFET logic die hybrid-bonded onto a NAND memory-wafer. Inset (red box) shows a cross-section through one of the hybrid-bonds connecting the logic die and memory wafer. Yellow arrows indicate the die-wafer interface.}
\label{fig:concept}
\end{figure}

The fabricated composite vehicle has layers added to a conventional 14nm foundry wafer and a custom NAND Flash memory wafer. Redistribution layers on the memory wafer support a combined pad-cage that can service both the logic die and the memory wafer (Fig.~\ref{fig:concept}). The top surface of both wafers is filled with an exposed pattern of few-$\mu$m square copper patches on a continuous square grid, with complementary patterns in the corners for alignment. Although Wafer-on-Wafer hybrid-bonding is highly mature and leads to good yield, Chip-on-Wafer hybrid-bonding allows proof-of-principle demonstration, avoiding the need to exactly match reticle pitch between logic and memory wafers at the prototype stage. Wafer-thinning and dicing are carefully optimized to improve overall yield and ensure testable prototypes. After bonding, composite vehicles are screened for NAND array health, via-chain health (threaded through $\sim$330 hybrid-bond pairs), and pad-continuity health. Suitable vehicles are then diced and wire-bonded.

\begin{figure}[tb]
\centering
\includegraphics[width=\columnwidth]{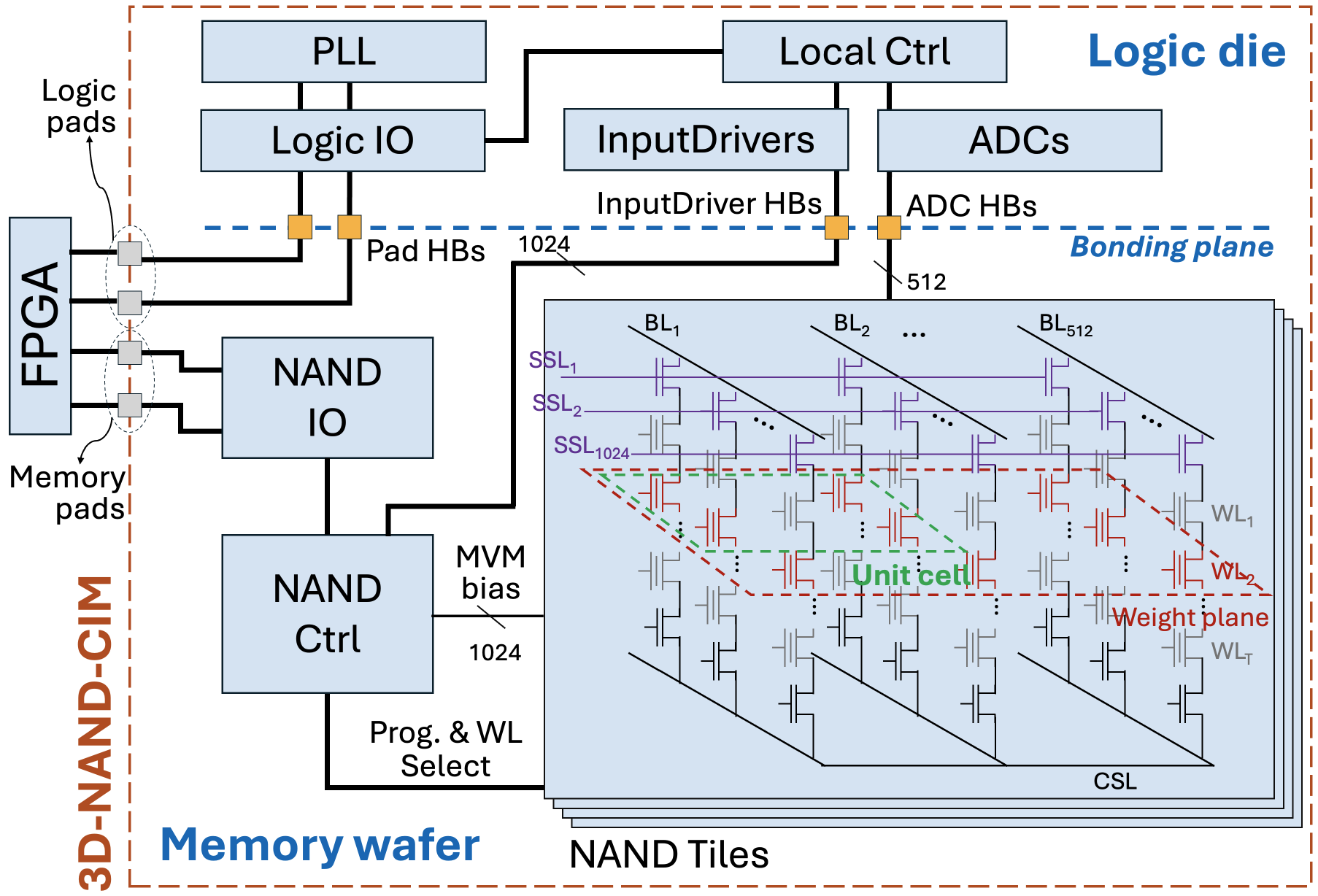}
\caption{3D-NAND-CIM operation: NAND programming is first performed using the memory wafer. Matrix-Vector Multiply (MVM) operations are controlled by the logic die, with input-drivers controlling memory row-lines, and ADCs measuring current along column-lines.}
\label{fig:operation}
\end{figure}

Testing comprises two steps: NAND device programming and Matrix-Vector-Multiplication (MVM) operations (Fig.~\ref{fig:operation}). The former is performed by performing block erase followed by careful device-by-device programming.
Programming voltage suitable for Fowler-Nordheim tunneling is applied to one of the wordline layers, while unselected wordlines are inhibited from programming.
As opposed to digital states detected by threshold-voltage, here the NAND devices are programmed to different analog-conductance states, using a 4-bit target (16 levels). After programming, the memory wafer is switched to a mode that biases a selected tier of a selected tile for MVM operation using the 1024 hybrid bonds carrying the input signals from the logic die, controlling the row-lines of the NAND array. Column-line currents from the operation are sensed by 512 ADCs in the logic die via a different set of hybrid bonds. Relevant hybrid-bonded signals are disconnected during programming to protect the logic die from the high voltages needed for NAND programming. Each NAND array has its own unique 512 column-line hybrid-bonds, so that each column-line is routed to one of 2048 ADCs on the logic die.
In this initial prototype vehicle, the four NAND arrays share the same 1024 row-line hybrid-bonds.

\begin{figure}[tb]
\centering
\includegraphics[width=\columnwidth]{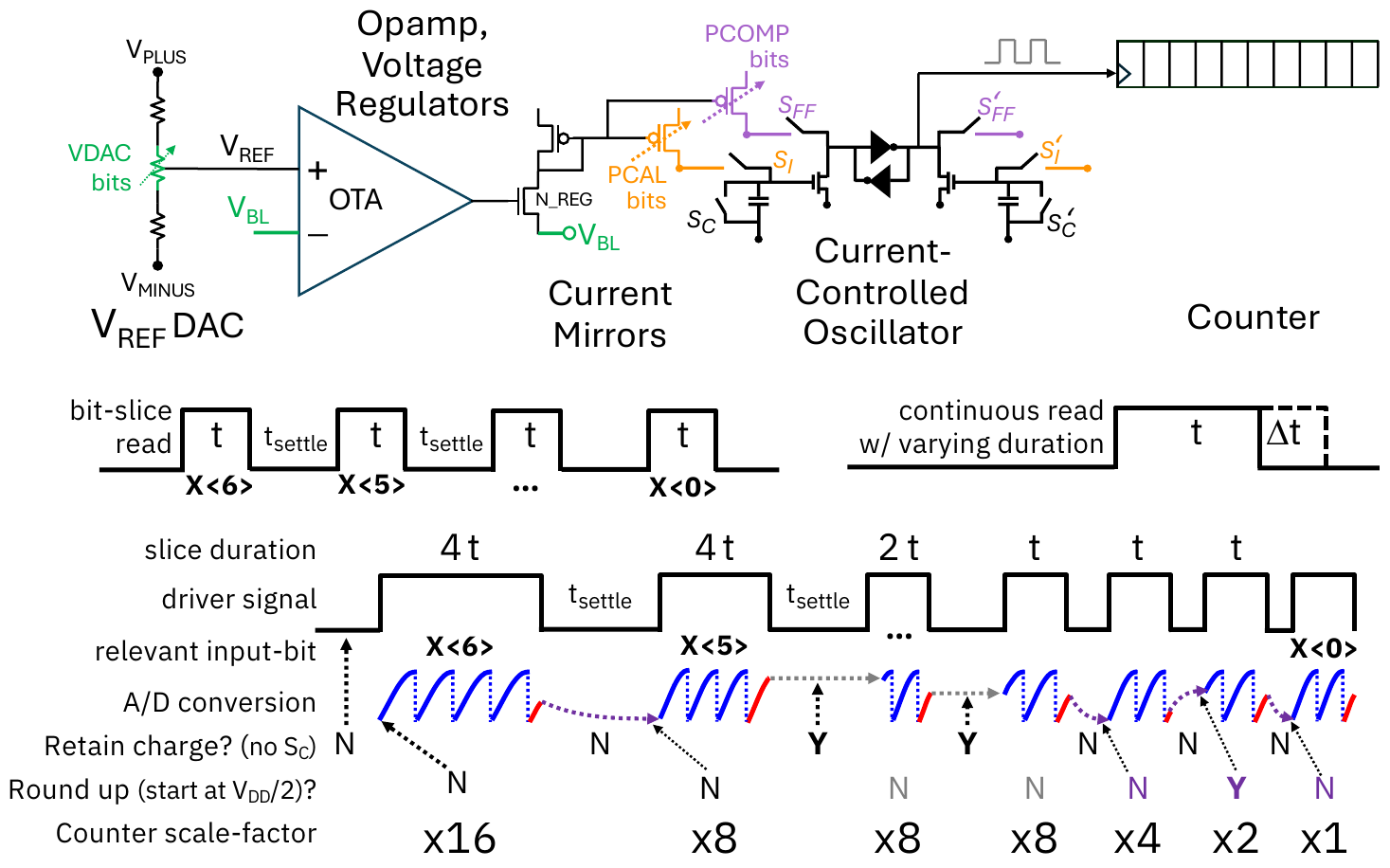}
\caption{(Top) Schematic of the ADC. (Middle and Bottom) Input significance is imposed either by pulse-duration, counter scale-factor, or pulse amplitude. Preserving capacitor charge between slices (switches $S_C,\:\!S^\prime_C$) or rounding reduces error.}
\label{fig:adc}
\end{figure}

\begin{figure}[tb]
\centering
\includegraphics[width=\columnwidth]{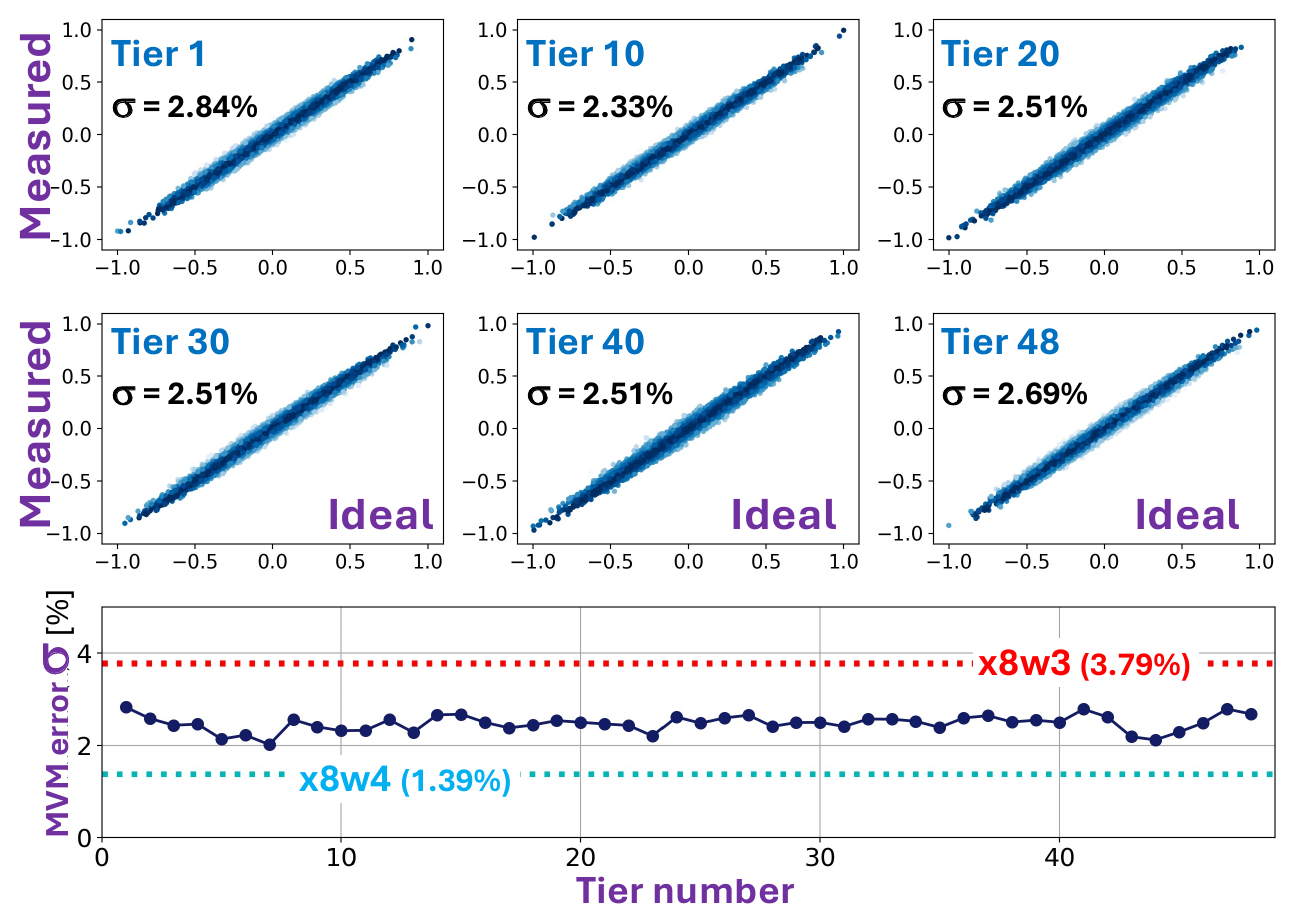}
\caption{(Top) Correlation plots showing measured vs.\ ideal output and MVM $\sigma$ for 6 of the first 48 tiers. (Bottom) Measured MVM-$\sigma$, for 256 weight-columns, for the first 48 tiers. Lines show quantization error of 8-bit digital inputs with 3- and 4-bit digital weights.}
\label{fig:tiers}
\end{figure}

Each MVM ``weight'' is encoded using a pair of physical row-lines and a pair of physical column-lines. The first physical row-line is driven in the case of a ``positive'' input-value; the second in the case of a ``negative'' input-value. Input-values are encoded by input-driver circuits on the logic die, which impose the right voltage and duration to implement a signed 8-bit input-value during MVM operation. A Local Controller (LC) on the logic die, clocked at $\sim$700MHz with an on-die PLL, allows loading of 8-bit input-values into a large SRAM, and then provides digital control over the input-driver circuits during MVM operation. To implement a signed weight, only two of the four NAND devices are programmed to the desired conductance, with the other two programmed to the deepest possible low-conductance state. The logical MVM value is thus computed by taking the difference between the integrated currents measured off the two physical column-lines, enabling a full signed-input times signed-weight operation. That said, use of ReLU functions (such as within the WideResNet-16-8-CNN implemented below) implies that subsequent inputs will be positive-only, allowing one weight per row-line and doubling the number of available rows per tile.

\begin{figure}[tb]
\centering
\includegraphics[width=\columnwidth]{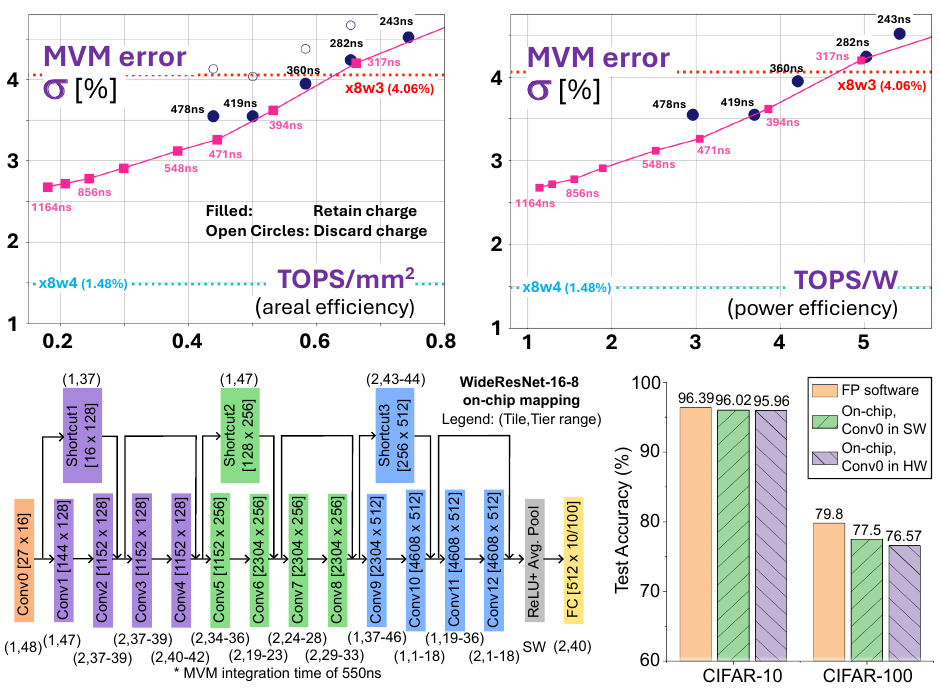}
\caption{(Top) Measured dependence of MVM~$\boldsymbol{\sigma}$ on TOPS/mm$^2$ and TOPS/W by decreasing total MVM time, using two sets of counter scale-factors (squares/circles). Charge-retention reduces errors. (Bottom) CIFAR-10/100 accuracy demonstrated on WideResNet-16-8 CNN.}
\label{fig:efficiency}
\end{figure}

Each physical column-line is connected to a dedicated CCO (Current-Controlled Oscillator)-based ADC (Fig.~\ref{fig:adc})\cite{HERMES_paper:2022}, each measuring the integrated current of its physical column-line during MVM operations. Up to 2048 on-chip ADCs measure MAC currents by having an analog front-end pass current to one of two small capacitors, followed by a digital back-end responsible for counting the number of times these capacitors are filled and emptied. ADCs include calibration capabilities for the VDAC holding the column-line voltage at a nominal read-voltage of 500mV, to control current-mirror gains, and for the quadratic correction needed to compensate for inherent charge-loss as the oscillation frequency increases\cite{HERMES_paper:2022}.

To implement a multi-bit input, MVM integration is broken into multiple ``bit-slices,'' either reducing the duration by 2$\times$, reducing the digital-significance by 2$\times$, or adjusting voltage amplitude from most to least significant bit. Digital-significance is imposed by changing which digital counter-bit is incremented each time a capacitor is filled. However, any residual charge accumulated on the active capacitor at the end of a bit-slice represents a partial-count which is unavoidably lost, representing an inherent quantization error. The ADC includes two special modes: the first allows this residual charge to be retained, for instance when duration will be used to halve the importance of the next bit-slice (e.g., the same counter-bit will be affected). The charge can still be retained even when digital-significance is halved.
If the charge is not retained, then the ADC can selectively allow the next bit-slice to start with the capacitor at 0V (``discard all charge'') or at V$_{DD}$/2 (``round up'').
Thus the sequence of duration, counter-bit significance, charge retention and rounding must be chosen to balance total integration duration, signal strength, and the quantization error due to uncounted charge. Control over the digital signaling for this MVM operation is provided by the highly-flexible LC. After integration, the digitized results are passed to an SRAM so that the FPGA can read the data out for multiple integrations.

\begin{figure}[h!]
\centering
\includegraphics[width=\columnwidth]{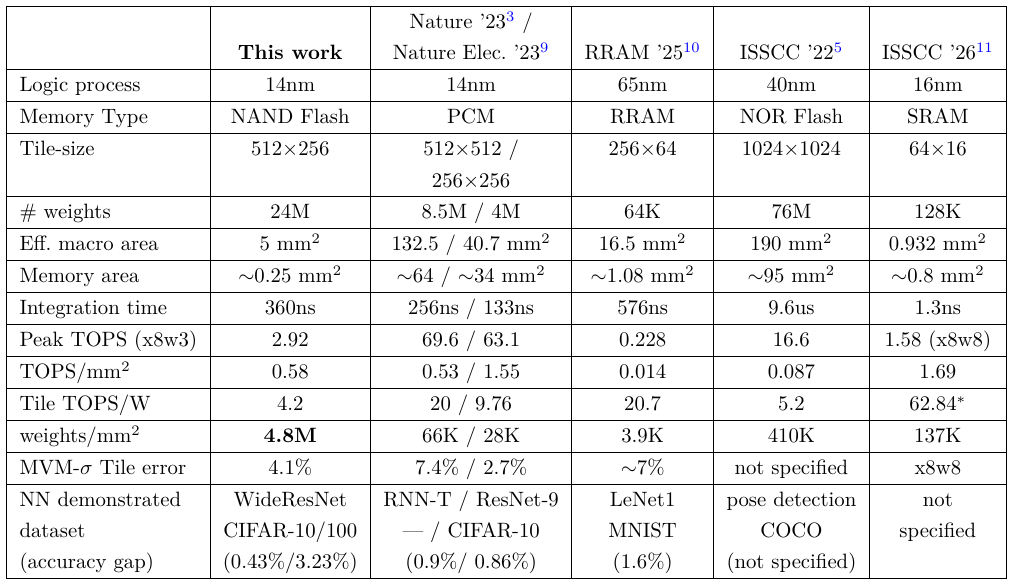}
\caption{Comparison table of recent NVM-based AIMC or SRAM-based IMC demos showing performance at 8-bit activations ($^\ast$not including the necessary but periodic weight-reloading of any SRAM-based macro). 1K = 1024, 1M = 1024$\times$1024.}
\label{fig:comparison}
\end{figure}

Figure~\ref{fig:tiers} (top) shows the results of MVM operations on 6 different tiers for one of the four Tiles, each plotting the 256 measured outputs as a function of the expected output. Here the input-vectors and weight matrices are taken from a synthetic dataset with ``4-quadrants,'' meaning that both weights and input activations have positive and negative values. The sparsity is 50\% for weights and 90\% for activations. Synthetic datasets allow for exploration of performance baselines, informing real-network co-optimization approaches. MVM noise is characterized by the standard-deviation ($\sigma$) of the error between the measured and desired outputs, divided by the maximum desired-output magnitude that was used to scale the outputs to [$-1.0$, 1.0]. As can be seen in Fig.~\ref{fig:tiers} (bottom), the MVM noise is fairly constant across the first 48 tiers, corresponding to less noise than would result from 8-bit digital inputs multiplied by 3-bit digital weights (``x8w3''), although not quite as low noise as would be achieved by ``x8w4.'' Note that MVM-$\sigma$ is a function of input and weight distributions and includes all kinds of non-idealities, such as IR-drop, weight-programming error, and quantization effects.

Figure~\ref{fig:efficiency} (top) shows the measured MVM-$\sigma$ value as a function of both energy-efficiency (TOPS/W) and areal-efficiency (TOPS/mm$^2$) as the total MVM time, including setup periods between the bit-slices, scales down. Preserving charge after every bit-slice, or even only on the bit-slices with identical counter scale-factor, had a marked benefit, but blindly rounding up between bit-slices had negligible impact.
A graceful tradeoff between the $\sigma$ value and energy/area efficiency is observed. With a fairly short MVM time of 360ns, an energy efficiency of 4.2 TOPS/W and area efficiency of 0.58 TOPS/mm$^2$ is achieved at a MVM error equivalent to x8w3 (3-bit) weight precision. Power is measured by performing a handful of different MVM operations in an infinite loop; area is estimated by the area needed for ADC, input-driver, SRAM and LC circuits (5mm$^2$). This area is significantly smaller than the test vehicle used here (enlarged by circuit and pad-cage reuse), but significantly larger than the actual area of the NAND Flash devices. In a future implementation, the SRAMs would be replaced by much slimmer double-buffers connected by communication infrastructure such as a 2-D mesh\cite{TVLSI_paper:2023}. Unnecessary features (bi-directionality, extra calibration circuits, and unused operating modes) would be removed from the ADCs and input-driver circuits, and the highly-flexible LC with its large SRAM program-buffer would get replaced by a Finite-State Machine significantly increasing areal efficiency.

To demonstrate that the achieved MVM error is sufficiently low to support high-accuracy during DNN inference, we programmed the weights of a WideResNet-16-8 Convolutional Neural Network (CNN, Fig.~\ref{fig:efficiency}) consisting of three residual groups (17 layers) with progressively increasing channel widths and isometry-promoting residual blocks with ReLU activations on the chip. The network trained on the CIFAR-10 (-100) dataset has 11~M weights all deployed on the chip. When executing all MVM operations of the weight layers deployed on the NAND tiers on-chip during inference, and the remaining operations in software, we achieved a test accuracy of 95.96\% (76.57\%), compared to the software baseline of 96.39\% (79.8\%).

\begin{figure}[h!]
\centering
\includegraphics[width=\columnwidth]{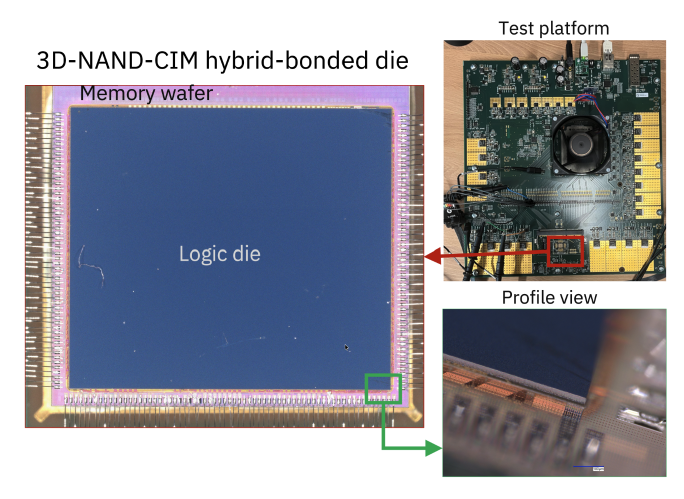}
\caption{Photo of hybrid-bonded vehicle, with logic die (blue) just inside memory wafer pad-cage (pink). Test platform hosts wire-bonded vehicle-under-test and connects to FPGA (not shown); a profile-view photo shows the corner of the vehicle-under-test.}
\label{fig:photo}
\end{figure}

The comparison table in Fig.~\ref{fig:comparison} shows that the present work is similar to the TOPS/W and TOPS/mm$^2$ of earlier NVM-based AIMC studies, while breaking new ground in weights/mm$^2$ by orders of magnitude, with significant improvement in sight with logic chip co-design. This massive change in the capacity-ceiling of AIMC can enable much larger models, from the edge to the cloud. Figure~\ref{fig:photo} shows a photo of the hybrid-bonded vehicle-pair, showing the back of the logic die surrounded on three sides by the pad-cage on the memory wafer, as well as the test platform and a photo showing the corner of the logic die sitting just inside the pad-cage on top of the memory wafer.

\textbf{Acknowledgements:}
This work was supported by the Defense Advanced Research Projects Agency (DARPA) under the OPTIMA program. The views, opinions and/or findings expressed are those of the authors and should not be interpreted as representing the official views or policies of DARPA or the U.S. Government. We thank the IBM Research AI HW center and Micron management for project support, IBM~Albany~Nanotech~Center for device and module fabrication, R.~Lockwood, P.~Solomon, C.~Osborn, P.~A.~Francese and I.~Boybat at IBM for technical support, and V.~Mukherjee, S.~Yamamichi, M.~Khare, H.~Bu and P.~Nagpurkar at IBM for management support.

% =========================================
%  REFERENCES  (Nature style)
% =========================================

\end{document}